\documentclass[showpacs,twocolumn,preprintnumbers,superscriptaddress,amsmath,amssymb,nofootinbib]{revtex4}
\usepackage{graphicx,color}
\usepackage[section]{placeins}
\newcommand{\cd}{\makebox[0.08cm]{$\cdot$}}

\newcommand{\sla}{\not\!}
\def\RR{\hbox{\it I\hskip -2.pt R }}
\begin{document}
\vspace{0.5cm}
\pagestyle{headings}

\title{Taylor-Lagrange renormalization scheme. \\ Application to light-front dynamics}

\author{P. Grang\'e}
\affiliation{Laboratoire de Physique Th\'eorique et Astroparticules,\\ Universit\'e Montpellier II, CNRS/IN2P3, Place E. Bataillon\\ F-34095 Montpellier Cedex 05, France}
\author{J.-F.~Mathiot}
\affiliation {Clermont Universit\'e, Universit\'e Blaise Pascal, Laboratoire de Physique
Corpusculaire, \\ BP10448, F-63000 Clermont-Ferrand, France}
\affiliation {CNRS/IN2P3, UMR 6533, LPC, F-63177 Aubi\`ere Cedex,
France}
\author{B. Mutet}
\affiliation{Laboratoire de Physique Th\'eorique et Astroparticules,\\ Universit\'e Montpellier II, CNRS/IN2P3, Place E. Bataillon\\ F-34095 Montpellier Cedex 05, France}
\author{E. Werner}
\affiliation{Institut f$\ddot u$r Theoretische Physik, Universit$\ddot a$t Regensburg,\\ Universit$\ddot a$tstrasse 31, \ D-93053 Regensburg,  Germany}
\bibliographystyle{unsrt}

\begin{abstract}
The  recently proposed renormalization scheme based on the definition of field operators as operator valued distributions acting 
on specific test functions is shown to be very convenient in explicit calculations of physical observables within the framework of  
light-front dynamics. We first recall the main properties of this 
procedure based on identities relating the test functions to their Taylor remainder of any order expressed in terms of Lagrange's 
formulas, hence the name given to this scheme. We thus show how it naturally applies to the calculation of state vectors of physical systems 
in the covariant formulation of light-front dynamics. As an example, we consider the case of the Yukawa model in the simple two-body Fock state truncation.
\end{abstract}
\pacs {11.10.Ef,11.10.-z,11.10.Gh,11.15.Tk\\
PCCF RI 09-04}
\maketitle

\section{Introduction}
The standard regularization and renormalization procedures are very well documented for any perturbative calculation of physical 
observables. They rely however on identification and cancellation of divergences by appropriate counterterms. While these procedures 
can be done in principle to any order in perturbation theory, it becomes particularly difficult, and cumbersome, to implement them beyond leading orders.  

We shall advocate in the following study the use of a general regularization/renormalization scheme based on the intrinsic properties of quantum field operators.
Indeed, the fundamental objects of relativistic field theory are covariant field
 operators recognized long ago as operator valued distributions (OPVD). These distributions are defined on test functions with well-defined properties \cite{BWH}. 
 The formulation of the $S$-matrix, well documented in \cite{Stu,Bolo}, was developed in the context
of OPVD in the 1970s by Epstein and Glaser \cite{EG,Sto} and more recently by Scharf \cite{Scharf}. 
Moreover, when causality conditions are
imposed on test functions,   $S$-matrix amplitudes, defined as  time ordered 
products of OPVD, are split into causal advanced and retarded pieces, in such a way that any singular 
behavior at equal space-time points is avoided \cite{aste}. More recently, the connection of this approach with the well known BPHZ 
renormalization scheme has been shown in \cite{GB}. 

We show in this study  how this new regularization/renormalization scheme can be applied in practice to physical systems in a nonperturbative framework. This scheme is quite general, and applies to any formulation of quantum field theory. As we shall see in the following, it is very well suited when relativistic bound state systems are formulated within light-front dynamics (LFD) \cite{brodsky} and its covariant formulation (CLFD) \cite{k82, cdkm}. In order to have a coherent presentation, we shall briefly recall the general properties of this approach, in the spirit of performing practical calculations.\\

The formulation of quantum fields as OPVD has recently been revisited by Grang\'e and Werner (GW)\cite{GW}.
Requiring that the resulting field theory should\\ 
\hspace*{0.5cm} $i)$   be independent of the form of the test functions used 
for the construction of the physical fields, and \\
\hspace*{0.5cm} $ii)$   preserve the basic Poincar\'e and Lorentz
 invariances,\\ GW are led to introduce test
functions in the form of partitions of unity (PU)  \cite{schwartz,PU} (see Appendix \ref{PU} for the definition of a PU).  With 
this choice of test functions, the physical fields are mathematically  well defined and show a familiar canonical form. 

By the very nature of the PU construction,
an arbitrary scale, characterising the building blocks of the decomposition of unity, comes naturally into the picture. Such a scale is 
the corner stone permitting the use of extended distributions in renormalization group studies.  

Moreover, the test functions are chosen as super regular test functions (SRTFs), i.e. they are  functions of finite extension - or finite support - vanishing with all their derivatives at their boundaries, either in the ultraviolet (UV) or in the infrared (IR) domains. The ultimate goal of the use of a PU-SRTF is to extend its support to infinity (UV case) or to zero (IR case) in such
a way that any physical amplitude remains finite when the PU is equal to $1$ everywhere in the integration domain. 

For a renormalizable field theory - e.g. with only a finite number of primitive divergences to
be handled - the SRTF specification is not compulsory in principle since the test function, and only a finite number of its derivatives, should vanish at the singularities \cite{EG}.
However, the choice of SRTFs lead to a very convenient and systematic construction of the extension of singular distributions in the spirit of the analysis of 
Epstein and Glaser. 
From the logical point of view, the 
introduction of SRTFs which are also PUs has not the character of an arbitrary restriction in the choice of test functions but is a mathematical necessity,
 if one wants to satisfy the two basic conditions  $(i)$ and $(ii)$ mentioned above.
 
The key feature of any  SRTF is its equality  with its Taylor remainder of any order. This remainder, written in terms of Lagrange's 
integral formulas, permits in turn convenient mathematical operations leading to the extension of singular distributions. We shall therefore
 call this general regularization/renormalization scheme the Taylor-Lagrange renormalization scheme (TLRS). In nonperturbative
calculations,  singular contributions cannot be singled out {\it a priori}  and their treatment is a major issue. The OPVD approach offers the immediate 
advantage of working with finite and mathematically well defined amplitudes. 

 Using SRTFs, we can handle, in principle, divergences of any degree. In this sense, the TLRS satisfies Weinberg's statement that 
 "the so-called nonrenormalizable theories are actually just as renormalizable as renormalizable theories"  \cite{wein}. In both cases, all relevant amplitudes will be finite from the very
  beginning, but scale dependent. The scale independence of physical observables will however only be achieved if all relevant terms 
  in the effective Lagrangian is considered. The number of such terms is finite in renormalizable theories, but infinite in 
  nonrenormalisable ones. In a given kinematical domain, one may however expect that only a limited number of such terms is dominant, 
  leading to the predictive power of effective field theories. In that case, Weinberg's statement is fully satisfied.

 Following the general properties of LFD, the state vector of any bound state system can be decomposed in Fock sectors.
 For practical calculations however, this Fock decomposition is truncated to a given order $N$, where $N$ is the maximal number of particles
 in the Fock sectors. Each Fock sector is then described by a many-body wave function, called the Fock component. 
As we shall see in Sec.~\ref{CLFD}, the construction of these many-body Fock components can be extended, in a very natural way, in order 
to include the test functions necessary to the regularization/renormalization of the physical amplitudes, as mentioned above. This leads 
to a very transparent and general formulation of LFD, with all amplitudes being finite but dependent on an arbitrary scale. 

Because of this intrinsic scale, we can do a renormalization group (RG)
analysis of our result, provided our renormalization scheme is well adapted to a Fock state truncation, and the Fock state
decomposition itself is convergent. As compared to the Pauli-Villars (PV) regularization scheme widely used in LFD calculations, the
regularization procedure inherent to the TLRS has several advantages. First of all, we consider in the construction  of the state vector only physical states, and
we do not have to enlarge the Fock space to deal with (negatively normed) PV fields, with one, two or more PV fields depending on the type 
of singularities.  Moreover, all amplitudes are finite from the very beginning, and numerical calculations should thus be easier
to control.

The plan of this article is the following. We recall in Sec. \ref{general} the main properties of the Taylor-Lagrange renormalization scheme. We then apply this scheme to light front dynamics in Sec. \ref{CLFD}. Perspectives are presented in Sec. \ref{pers}.
We  detailed in Appendix \ref{PU} the construction of the test function. The explicit calculation of the Fock components in CLFD is done
in Appendix \ref{selfe}. 

\section{Taylor-Lagrange renormalization scheme} \label{general}
We follow in this section  very closely the developments made in Ref.~Ê\cite{GW}. We start, for simplicity, from the general solution of 
the Klein-Gordon equation. It is given by a distribution, $\phi$, which defines a functional, $\Phi$, with respect to a test function 
$\rho$ \cite{schwartz} according to (in $D$ dimensions)
\begin{equation}
\Phi(\rho) \equiv < \phi,\rho > = \int d^Dy \phi(y) \rho(y)\ .
\end{equation}
The physical field $\varphi(x)$ is then defined in terms of the translation, $T_x$,  of $\Phi(\rho)$, which, in flat space, is given 
by
\begin{equation} \label{conv}
\varphi(x) \equiv T_x \Phi(\rho) =\int d^Dy \phi(y) \rho(x-y)\ .
\end{equation}
Because of  the properties of partial integrations with Schwartz test functions, $\varphi(x)$ obeys also the original field 
equation, Klein-Gordon in this case. 

Since the test function is a symmetric function of $x-y$, its Fourier transform  writes
\begin{equation}
\rho(x-y) = \int \frac{d^Dq}{(2\pi)^4} e^{iq.(x-y)} f(q_0^2, {\bf q}^2)\ .
\end{equation}
The  field $\phi(x)$ can thus be decomposed, in momentum space, in terms of creation and destruction operators. After integration over $p_0$, 
we get
\begin{equation}
\!\varphi (x)\!=\!\!\int\!\frac{d^{D-1}{\bf p}}{(2\pi)^3}\frac{f(\varepsilon_p^2,{\bf p}^2)}{2\varepsilon_p}
\left[a^+_{\bf p} e^{i{\bf p.x}}+a_{\bf p}e^{-i{\bf p.x}}\right],\ 
\end{equation}
with  $\varepsilon^2_p = {\bf p}^2+m^2$.  Clearly, to have a procedure independent of the choice of test functions, the convolution of 
$\varphi(x)$ itself according to (\ref{conv}) should leave  $\varphi(x)$ unchanged \cite{GW}. We shall come back to this point later on.  

It is apparent that test functions should be attached to each fermion and boson fields, when deriving the effective Hamiltonian. Each 
propagator being the contraction of two fields should be proportional to $f^2$. In order to have a dimensionless argument for f, 
we shall introduce an arbitrary scale $\Lambda$ to "measure" all momenta. $\Lambda$ can be any of the masses of the constituents. In order to deal with massless theories, we shall take some arbitrary value. The final expression of any amplitude 
should be independent of $\Lambda$. In CLFD, the test function is thus a function of $\frac{{\bf p}^2}{\Lambda^2}$ only. One can already
notice, at this stage, that there is no difference between   $\Lambda$ and  $\eta \Lambda$, with $\eta$ a positive number, and the test
function should embody this important scaling information.

\subsection{Singular distributions, Taylor's remainder and  Lagrange's formulas} \label{lagrange}
Let us assume that $T(X)$ is a distribution with a singularity of order $k$ at the origin of $\RR^{\! D}$, where $k$ is defined by
\begin{equation*}
k=inf \{s:{\displaystyle \lim_{\lambda \rightarrow 0}}\lambda ^{s} T(\lambda X) = 0\} - D\ .
\end{equation*}
To give an example, let us consider, for $D=1$, the distribution $T(X)$ behaving like $\frac{1}{X^l}$ with $l=1,2,3 ...$ when  
$X \rightarrow 0$ but with faster decrease than $\frac{1}{X}$ at infinity. We have $k=l+1-1=l$. Any physical amplitude is written in a schematic way as
\begin{equation} \label{ampli1}
{\cal A}=\int_0^\infty dX\ T(X)\  f(X)\ .
\end{equation}
If the test function $f(X)$ does not tend to zero fast enough at the origin, the classical mathematical method to give a meaning to 
${\cal A}$  is to perform a Taylor expansion of $f$ around $X=0$ and to suppress from $f$ as many terms in the series as is necessary to 
obtain a finite contribution from the region around $X=0$. The resulting function is just the Taylor remainder at $X=0$, denoted by $R^k(f)$, and defined by
\begin{equation}
R^k(f)(X)=f(X)-\sum_{n=0}^k \frac{X^n}{n!} \partial^n f(X) \vert_{X=0}\ .
\end{equation}
This method has been
known in functional analysis for a century and leads to the concept of pseudofunctions \cite{schwartz}. In the diverging functional 
${\cal A}$, $f(X)$ is  thus replaced by its Taylor remainder $R^{k}(f)(X)$ of order $k$. It is also at the heart of the BPHZ renormalization scheme \cite{bphz}.

However, when $f(X)$ is a SRTF vanishing at the origin
with all its derivatives, $f$ is strictly equal to its Taylor remainder of any order. In that case, the physical amplitude obeys the identity
\begin{equation} \label{ampli2}
{\cal A}=\int_0^\infty dX \ T(X)f(X)= \int_0^\infty  dX \ T(X) R^{k}(f)(X)  \ .
\end{equation}
In order to define the extension, denoted by $\widetilde{T}(X)$, of the distribution $T(X)$  over the entire space, i.e. in the limit where $f \to 1$, we can use Lagrange's formula for the Taylor remainder, hence our denomination Taylor-Lagrange renormalization scheme. Their use was first advocated by Brunetti and Fredenhagen \cite{BF}. The explicit procedure, which involves a transposition of $R^{k}(f)(X) $ to $T(X)$, is detailed in the following. It leads to
\begin{equation} \label{IRA1}
{\cal A}={\displaystyle \lim_{f \rightarrow 1}} \int_0^\infty dX\ \widetilde{T}(X)\ f(X)\ .
\end{equation}
The limit $f \to 1$ can now be done safely since the integral of $\widetilde T$ over $X$ is finite. Note that the functional $\widetilde{T}$ is defined
modulo a sum of the distribution $\delta(X)$ and its derivatives of order less or equal than  $k$.
This sum  is killed  by $f$ in the functional, and should therefore not be considered in (\ref{IRA1}).  This is however not the case for the expression of $\widetilde{T}(X)$  considered in isolation of the
integral. A similar analysis can be done when $X \to \infty$, i.e. in the UV domain. 

Lagrange's formula can take several forms depending on which 
kinematical domain, UV or IR, we are interested in. We shall separate in the following the test function used in the UV domain, $f^>$, from the one used in the IR domain, $f^<$, in order to separate their different scaling behavior. In the UV domain, we have for instance
\begin{equation}
f^>(X)=-\int_1^\infty dt \ \partial_t \left[ f^>(Xt) \right]\ ,
\end{equation}
for the simplest case where the function $f^>$  alone is zero at infinity. 
One may rewrite this identity in a slightly different form, by shifting the derivative on $t$ to a derivative on $X$. We get
\begin{equation}
f^>(X)=-X\int_1^\infty \frac{dt}{t} \partial_X \left[ f^>(Xt) \right]\ .
\label{La2}
\end{equation}
Note that the Lagrange formula can be written, in the case of a function of $X^2$:
\begin{equation}
f^>(X^2)=-X\int_1^\infty \frac{dt}{t} \partial_X \left[ f^>(X^2t^2) \right]\ .
\label{La22}
\end{equation}
 It can be generalized if the function $f^>$ is a SRTF, leading to 
 \begin{equation} \label{la3}
f^>(X)\!=\!-\frac{X}{k!}\!\!\int_1^\infty\! \frac{dt}{t} (1-t)^k \partial_X^{k+1} \left[ X^k f^>(Xt) \right],
\end{equation}
for any integer $k \ge 0$. 

Similarly, one can write in the IR domain:
 \begin{equation} \label{la3IR}
f^<(X)=\frac{X}{k!}\int_0^1 \frac{dt}{t} (1-t)^k \partial_X^{k+1} \left[X^k  f^<(Xt) \right].
\end{equation} 
Lagrange's formula can be written in  different ways since, if $f(X)$ is a SRTF, then $X^k f(X)$ is also a SRTF
 for any value of $k$. So that (\ref{la3IR}) can be rewritten - with the replacement $X^k f(X) \to f(X)$ - like
\begin{equation} \label{la33}
f^<(X)\!=\!\frac{X^{k+1}}{k!}\!\int_0^1\!\! \frac{dt}{t^{k+1}} (1-t)^k \partial_X^{k+1} \left[f^<(Xt) \right],
\end{equation}
which shows explicitly a zero of order $k+1$ at the origin, as mentioned earlier.

Finally, under the scale transformation $X\to aX$, the Lagrange formula writes, after the change of variable  $t \to a t$:
\begin{equation} \label{la3a}
f^>(aX)=-\frac{X}{a^k k!}\int_a^\infty\frac{dt}{t} (a-t)^k \partial_X^{k+1}\!\left[ X^k f^>(Xt)\right].
\end{equation}

\subsection{Evaluation of amplitudes with test functions} \label{extension}
 From now on, we shall work with super regular test functions chosen among the partitions of unity. The property that the integral (\ref{ampli1})
 is independent of the precise choice of this partition of unity will come out explicitly \cite{schwartz,PU}.  Moreover, if $f$ is a PU with 
 a given support, any power of $f$, $f^n$, is also a PU with the same support. They are said to be equivalent. In the limit where $f \to 1$ 
 over the whole space, they are strictly equal.  This property is essential in order to have in (\ref{conv}) a field independent  of the
 construction of $f$ and preserve the basic Poincar\'e and Lorentz invariances \cite{GW}. One possible realization of a SRTF partition 
 of unity is detailed in Appendix \ref{PU}. By construction, the test function is $1$ everywhere except in the vicinity of the boundaries. 
 
 To begin with, we shall concentrate in this section on distributions singular in the UV domain. 
 The upper boundary - defining the extension of the support of $f$ -  is taken to be $H$ in the UV domain, so that $X \le H$.  We shall 
 denote by $f(X;H)$ the test function in order to keep track of this boundary.
 
We suppose, in a first step, that the UV divergence is of a logarithmic type. This corresponds to the lowest order singularity with $k=0$. 
Our formalism will be easily  generalized to distributions with higher singularities. Using (\ref{La2}), we have
\begin{equation}
\!\!{\cal A}\!=\!-\!\int_0^\infty \!\! dX T(X) X\int_1^\infty \frac{dt}{t} \partial_X \left[ f^>(Xt;H) \right].
\end{equation}
The integration limits are determined by the support of the test function. We thus have
\begin{equation}
\!\!{\cal A}\!=\!-\!\int_0^{H} \!\!dX T(X) X\int_1^{\frac{H}{X}}\frac{dt}{t} \partial_X \left[ f^>(Xt;H) \right]\ .
\label{Ain}
\end{equation}
By integrating (\ref{Ain}) by part on $X$, we  have
\begin{eqnarray} \label{Afi}
\!\!{\cal A}&=&-X T(X) \left.\int_1^{\frac{H}{X}} \frac{dt}{t} f^>(Xt;H) \right|_{X=0}^{X=H} \nonumber \\
&+&\!\!\int_0^{H}\!\! dX \partial_X \left[ X T(X)\right] \int_1^{\frac{H}{X}} \frac{dt}{t} f^>(Xt;H) \nonumber \\
&+&\!\!\int_0^{H}\!\! dX X T(X) \left( \frac{H}{X}\right) ^\prime \left.\frac{f^>(Xt;H)}{t}\right|_{t=\frac{H}{X}}. 
\end{eqnarray}
Because of the boundary conditions of the test function, the first and third terms in the right-hand side of Eq.(\ref{Afi}) are equal to zero,
and it remains
\begin{equation}
{\cal A}=\int_0^{H} dX\  \partial_X\  \left[ X T(X)\right] \int_1^{\frac{H}{X}} \frac{dt}{t} f^>(Xt;H)\ .
\label{afin}
\end{equation}
Up to now, we have only made use of identities, and of the fact that the test function should be zero at some boundary defined by $X_{max}=H$. 
Calculating (\ref{afin}) with a constant $H$  leads to the usual result using ordinary cut-off, i.e. an amplitude which behaves like $Log[H]$. 
The limit $f\to 1$ would thus be achieved by letting $H \to \infty$, leading to a logarithmic divergence. This is equivalent to 
the standard procedure using a sharp cut-off on the variable $X$. Clearly  one should consider another construction of the test function in the
 UV - as well as IR - keeping track, in the ultimate limiting process, of the generic scaling properties embodied in any PU.

\subsection{Test functions with running support} \label{DTF}
 In order to go beyond this naive, infinite $H$-limit, we shall consider a boundary condition for which the boundary $H$ depends on the 
 running variable $X$ \cite{GW}. We thus define a function $g(X)$ by:
\begin{equation} \label{running}
H(X)\equiv \eta^2Xg(X)\ ,
\end{equation}
up to an additive arbitrary finite constant irrelevant in the UV domain. This function depends on an arbitrary  dimensionless scale $\eta^2$ which is related to
the shape of the test function near $H$. In the next subsection we detail the conditions to be imposed on $g(X)$.

The extension of the procedure explained in \ref{extension} to deal with running test functions should be done with care. Replacing directly
$H$ by $H(X)$ and $f^>(X;H)$ by a single function $F^>$ depending on X with $F^>(X)\equiv f^>(X;H(X))$ would not make any difference 
since $F^>(X)$ is a function which vanishes at the upper boundary denoted by $X_{max}$ and thus (\ref{afin}) will still behave like 
$Log[X_{max}]$, which is divergent when $X_{max} \to \infty$. 

Clearly, Lagrange's formula in (\ref{la3}) applies for a given $X$-value at {\it a fixed support.} So that in the $t$-integral, $X t$  
just moves  the argument of $f$ along this fixed support up to a value of $t$ such that $Xt=H(X)$. In order to keep track of this fixed
support, it is convenient to  consider the function $F^>$of the two variables $X$ and $Y$ defined by
\begin{equation} \label{fxy}
F^>(X,Y)\equiv f^>(X;H(Y))\ ,
\end{equation}
and the test function $f$ is now given by
\begin{equation}
f^>(X;H(X)) \equiv \left. F^>(X,Y) \right|_{Y=X}\ .
\end{equation}
Each specific Lagrange formula is then applied to the $X$ dependence of $F$ only. In the UV domain for instance, it leads to:
\begin{eqnarray}
 \!\!\left.F^>(X,Y)\right|_{Y=X}\!&=&\!-\!\int_1^{\frac{H(X)}{X}}\! dt \partial_t \left[ F^>(Xt,Y)\right]_{Y=X} \nonumber\\
\!\!\!&=&\!-\!X\!\int_1^{\frac{H(X)}{X}}\! \frac{dt}{t} \partial_X \left[ f^>(Xt,H(X))\right] \nonumber \\
+ X H'(X)\!& &\!\!\!\!\!\int_1^{\frac{H(X)}{X}}\! \frac{dt}{t} \left.\partial_Y F^>(Xt,Y)\right|_{Y=X}.
\label{La2d}
\end{eqnarray}
We shall construct the test function in such a way that the last term on the right hand side of (\ref{La2d}) vanishes because in the UV there
 will be no overlapping of the domain where  $\partial_Y F^>(Xt,Y)$ is finite with the domain of integration on $t$.

\subsection{Extension of singular distributions }\label{distri}
\subsubsection{In the ultraviolet domain}
Following the general procedure detailed in Sec. \ref{extension}, we can write, with (\ref{running}),(\ref{fxy}) and (\ref{La2d}) 
\begin{equation}
{\cal A}\!=\!\!\int_0^{X_{max}}\!\!\!\!\!\!\!\! dX\  \partial_X\ \!\!\! \left[ X T(X)\right] \! \int_1^{\eta^2 g(X)}\!\! \frac{dt}{t} 
F^>(Xt,X)\ .
\label{afind}
\end{equation}
In order to extend the test function to $1$ on the whole space, we shall consider a set of function $g(X)$, denoted by $g_\alpha(X)$, where 
by construction $\alpha$ is a real positive number less than $1$, and where the limit $\alpha \to 1^-$ corresponds to $f \to 1$. We note that,
 since $\partial_X \left[ X T(X)\right] $ will give by itself a finite $X$-integral when $X_{max} \to \infty$ 
 (the singularity of $T^>(X)$ has been chosen of order $k=0$),  the process of extending  the test function to unity  over the whole
  integration domain in $X$ may be chosen most simply to result, first, in an $X$-independent $t$-integral and, second, 
  to $X_{max} \to \infty$, where $X_{max}=H(X_{max})$, and vice versa.
  
Different choices of $g_\alpha(X)$ are possible, provided the above two properties are satisfied. Anticipating  the discussion 
of the next subsection, it should also provide a meaningful extension of singular distributions in the IR domain, as detailed in
  Appendix~\ref{dyna}. Following \cite{GW} we shall choose, as an example
\begin{equation} \label{galpha}
g_\alpha(X)= X^{(\alpha-1)}\ , 
\end{equation}
One might also consider other choices, like $g_\alpha(X)= 1+(\alpha-1) X$ or Exp$[(\alpha-1) X]$. The difference between these choices will
 mainly be in the rate at which $X_{max}$ goes to infinity when $\alpha$ approaches $1^-$, the extended distribution being the same, as 
 it should.

To see that the limit $\alpha \to 1^-$ corresponds also to the extension of the PU to the whole integration
domain, it is sufficient to look at the maximal value of $X$ at which the PU test function goes to zero. It is defined by
 $X_{max}=H(X_{max})$ so that, with (\ref{galpha})
\begin{equation}
X_{max}=(\eta^2)^{\frac{1}{1-\alpha}}\ ,
\end{equation}
The limit $X_{max} \to \infty$ is achieved by the condition $\alpha \to 1^-$, with $\eta^2>1$. The running support of the PU test function
stretches then over the whole integration domain.   

For finite values of $X$, $\lim_{\alpha \to 1^-}g_\alpha(X)=1$, so that the upper limit of integration on $t$ in (\ref{afind}) is just 
$\eta^2$. Near $X_{max}$, i.e. in the asymptotic region where the test function is less than $1$ and goes to zero ({\it cf} Appendix A2 and
Fig. 6), we have $g_\alpha(X) \simeq1/\eta^2$. This region of integration, however, does not contribute at all to the amplitude $\cal A$ 
since the integral on $X$ is now completely finite and thus insensitive to the limit of very large $X$, when $X \simeq X_{max} \to \infty$.

We can thus take in (\ref{afind}) the limit $\alpha \to 1^-$. The physical amplitude writes 
\begin{eqnarray}
{\cal A}&=&\int_0^\infty dX \ \partial_X \left[ X T(X)\right] \int_1^{\eta^2} \frac{dt}{t}\label{Ato} \nonumber \\
&\equiv&\int_0^\infty dX \ \widetilde T^>(X)\ ,
\end{eqnarray}
with 
\begin{equation}
\widetilde T^>(X)\equiv \partial_X \left[ X T(X)\right] \mbox{Log}(\eta^2) \ .
\end{equation}
The extension $\widetilde T^>(X)$ of the distribution $T(X)$ in the UV domain depends logarithmically on the arbitrary scale $\eta^2$, 
with $\eta^2>1$. The amplitude (\ref{Ato}) is now completely finite. 

The generalization of this procedure to singular distributions of any order can easily be done in a very similar way  \cite{GW}. It leads to the 
following extension
\begin{equation}
\!\!\widetilde T^>(X)\equiv\frac{(-X)^{k}}{k!} \partial_X^{k+1} \left[ X T(X)\right] \int_1^{\eta^2} \frac{dt}{t} (1-t)^k,\
\end{equation}
for $k\geq 0$, while $\widetilde T^>(X)=T(X)$ for $k < 0$. 

Note that we do not need the explicit form of the test function in the derivation of the extended distribution $\widetilde T^>(X)$. 
We only rely on its mathematical properties  and on its running construction. 

\subsubsection{In the infrared domain}
The extension of singular distributions in the IR domain has been studied in details in Ref. \cite{GB}. We recall here how it should
 be understood in terms of SRTFs, as proposed in \cite{GW}. In order to keep track of the explicit regularization  
 of the $X=0$ singularity by the test function, we can study the amplitude in terms of $[f^<]^2 \sim f^<$ and write 
\begin{equation}
{\cal A}=\int_0^\infty dX T(X) [f^<(X)]^2\ .
\end{equation}
We can thus apply the Lagrange formula (\ref{la3IR}) to one of the $f^<$'s only, and get
\begin{multline}
{\cal A}=\int_0^\infty dX T(X) f^<(X) \frac{X^{k+1}}{k!} \\
\left. \int_{0}^{1} dt \frac{(1-t)^k }{t^{k+1}}\partial_{X}^{k+1} F^<[Xt,Y] \right|_{Y=X}\ .
\end{multline}
By a change of variable $Xt \to X$ we have 
\begin{multline}
{\cal A}=\int_0^\infty dX  \frac{X^{k+1}}{k!} 
\int_{0}^{1} dt \ T\left[\frac{X}{t}\right] \\
f^<\left[\frac{X}{t}\right] \frac{(1-t)^k }{t^{k+2}}\partial_{X}^{k+1}\left.F^<[X,Y] 
\right|_{Y=\frac{X}{t}},
\end{multline}
Using the boundary condition on $f^<$ resulting from the behavior of the test function near the origin and  detailed in Appendix \ref{dyna}, 
we can write, after integration by part and in the limit where the test function extends to $1$ on the whole space\footnote{with, however,
the same considerations following (\ref{IRA1}).} 
\begin{equation} \label{IR}
{\cal A}={\displaystyle \lim_{f \rightarrow 1}} \int_0^\infty dX\ \widetilde{T}(X)\ f(X)\ ,
\end{equation}
with
\begin{multline} \label{TIR1}
\widetilde T^<(X) =(-1)^{k+1}\partial_{(X)}^{k+1} \left[ \frac{X^{k+1}}{k!} \right. \\
\left. \int_{\tilde \eta X}^{1} dt \frac{(1-t)^k}{t^{k+2} }T\left( \frac{X}{t}\right)\right]\ .
\end{multline}
Note that the derivatives in (\ref{TIR1}) and (\ref{TIR2})  below  have to be taken in the sense of distributions 
({\it cf} Appendix B3). The scale $\tilde \eta$ is positive and arbitrary.

For an homogeneous distribution, with $T[X/t]=t^k T(X)$, the $t$ integration can be carried 
out to give 
\begin{multline} \label{TIR2}
\widetilde T^<(X)=(-1)^{k}\partial_{X}^{k+1} \left[ \frac{X^{k+1}}{k!} T(X) \mbox{Log} (\tilde \eta X)\right] \\
	+ \frac{(-1)^k}{k!} H_k C^k \delta^{(k)}(X)\ ,
\end{multline}
with
\begin{eqnarray}
H_k&=&\sum_{p=1}^k \frac{(-1)^{p+1}}{p}{ k\choose p} = \gamma + \psi(k+1) \nonumber \\
C^k&=&\int_{(X=1)} T(X) X^k dS \nonumber
\end{eqnarray}
and $\psi$ is the usual digamma function with $\psi(1)=-\gamma$. 

The extension $\widetilde T^<(X)$ differs from the original distribution 
$T(X)$ only at the singularity. 
%
\section {Application to Light-Front dynamics} \label{CLFD}
\subsection{Covariant formulation of light-front dynamics}
In CLFD, the state vector of a physical system is defined on the LF plane of general
orientation  $\omega\cd x=\xi$, where $\omega$  is  an arbitrary
lightlike four-vector $\omega^2$=0, and $\xi$
is the LF "time" \cite{cdkm}. Standard light-front dynamics is recovered by choosing $\omega = (1,0,0,-1)$. We shall take $\xi=0$, for convenience.

Any bound system is entirely described by its state vector $\phi_\omega^{J\sigma}(p)$. It corresponds to definite values for the
mass $M$, the four-momentum $p$, and the total angular momentum
$J$ with projection $\sigma$ onto the $z$ axis in the rest frame,
and forms a representation of the Poincar\'e
group. It depends on the position $\omega$ of the light-front. The four-dimensional angular momentum
operator $\hat J$  is represented as a sum of the free and interaction parts:
\begin{equation}
\label{kt2} \hat{J}_{\rho\nu}=\hat{J}^{(0)}_{\rho\nu}
+\hat{J}^{int}_{\rho\nu}\ .
\end{equation}
In terms of the interaction Hamiltonian, we have
\begin{equation}\label{kt5}
 \hat{J}^{int}_{\rho\nu}=\int H^{int}(x)(x_{\rho}\omega_{\nu} -x_{\nu}
\omega_{\rho}) \delta(\omega\cd x)\ d^4x\ .
\end{equation}
From the general transformation properties of both the  state
vector and the LF plane, it follows~\cite{k82,cdkm} that
\begin{equation}\label{kt12}
\hat{J}^{int}_{\rho\nu} \ \phi_\omega^{J\sigma}(p)=
\hat{L}_{\rho\nu}(\omega)\phi_\omega^{J\sigma}(p) \ ,
\end{equation}
where
\begin{equation}\label{kt13}
\hat{L}_{\rho\nu}(\omega) =i\left(\omega_{\rho}
\frac{\partial}{\partial\omega^{\nu}} -\omega_{\nu}
\frac{\partial}{\partial\omega^{\rho}}\right)\ .
\end{equation}
The Eq.~(\ref{kt12}) is called the {\it angular condition}.

This equation does not contain the interaction Hamiltonian,
once  $\phi$ satisfies the Poincar\'e group equations. The construction of the wave functions of
states with definite total angular momentum becomes therefore {\it
a purely kinematical problem}. The dynamical dependence of the wave functions
on the LF plane orientation now turns into their explicit
dependence on the four-vector $\omega$. Such a
separation, in a covariant way, of kinematical and dynamical
transformations is a definite advantage of CLFD as compared to
standard LFD on the plane $t+z=0$.

According to the general properties of LFD, we decompose the state vector of a physical
system in Fock sectors. Schematically, we have
\begin{equation}
\phi^{J\sigma }_{\omega}(p) \equiv \vert 1 \rangle +
\vert 2 \rangle +
 \dots + \vert n \rangle + \dots
\end{equation}
Each term on the right-hand side denotes a state with a fixed number of
particles.  In
the Yukawa model, the analytical form of the Fock decomposition is
\begin{widetext}
\begin{eqnarray}
\phi^{J\sigma
}_{\omega}(p)&=&\sum_{n=1}^{\infty
}\frac{(2\pi)^{3/2}}{(n-1)!}\sum_{\sigma'} \int\phi_{n,\sigma\sigma'}(k_{1}\ldots
k_{n},p,\omega\tau_{n}) a^{\dag}_{\sigma'}({\bf
k}_{1})c^{\dag}({\bf k}_{2})\ldots c^{\dag}({\bf k}_{n})|0\rangle \nonumber \\
\label{eq21} &\times &\delta^{(4)}(k_{1}+\ldots
+k_{n}-p-\omega\tau_{n})2(\omega\cd p)d\tau_{n }\prod_{l
=1}^n \frac{d^3k_{l
}}{(2\pi)^{3/2}\sqrt{2\varepsilon_{k_{l }}}}\ ,
\label{twobody}
\end{eqnarray}
\end{widetext}
where $\phi_{n,\sigma\sigma'}(\ldots)$ is the $n$-body LF wave function, the so-called Fock component,
describing the state made of one free fermion and
$(n-1)$ free bosons, $a^{\dag}$ ($c^\dag$) are the free fermion
(boson) creation operators, $\varepsilon_{k_{l}}=\sqrt{{\bf
k}_{l}^{2}+m_{l}^{2}}$, and $m_{l}$ is the mass of the particle $l$
with the four-momentum $k_{l}$. The combinatorial factor $1/(n-1)!$
 is introduced in order to take into account the identity
of bosons. We do not consider in this approximation excitation of fermion-antifermion states ("quenched approximation").

The variable
$\tau_n$ describes how far off the energy shell the constituents
are. 
In practical calculations, the infinite sum over $n$ is
truncated by retaining terms with $n$ which does not exceed a
given number $N$, while those with $n>N$ are neglected.
Decompositions analogous to Eq.~(\ref{twobody}) can be easily
written for QED~\cite{kms_04} or for a purely scalar
system~\cite{bckm}.

The construction of the spin
structure of the wave functions $\phi_{n,\sigma \sigma'}$ is very
simple, since it is purely kinematical. This structure should incorporate
however $\omega$-dependent components in order to fulfill the angular
condition~(\ref{kt12}). It is convenient to decompose each wave
function $\phi_{n,\sigma \sigma'}$ into invariant amplitudes
constructed from the particle four-momenta (including the
four-vector $\omega$!) and spin structures (matrices, bispinors,
etc.). In the Yukawa model  we have for instance, for $N=2$,
\begin{subequations}
\label{Yuphi}
\begin{eqnarray}
\label{oneone}
\phi_{1,\sigma\sigma'}&= &\varphi_1\  \bar{u}_{\sigma'}(k_1)u_{\sigma}(p)\ ,\\
\phi_{2,\sigma\sigma'}&=&\bar{u}_{\sigma'}(k_1) \left[\varphi_2  +
\varphi_2^\omega\ \frac{m \sla \omega }{\omega \cd p}\right]
u_{\sigma}(p)\ , \label{onetwo}
\end{eqnarray}
\end{subequations}
since no other independent spin structures can be constructed.
Here $u$'s are free bispinors of constituent  mass $m$; $\varphi_1$, $\varphi_2$, and $\varphi^\omega_2$ are
scalar functions determined by the dynamics. 

\begin{figure}[btph] 
\begin{center} 
\includegraphics[width=15pc]{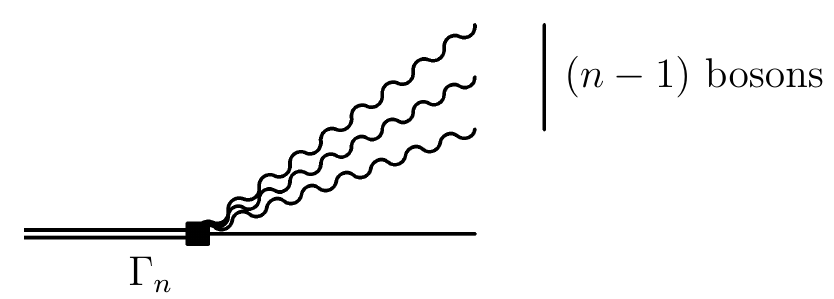}
\caption{Vertex function of order $n$. \label{gamman}}
\end{center}
\end{figure}

The eigenvalue equations for the Fock components can be obtained from
the Poincar\'e group equation  
\begin{equation}
\hat P^2 \phi(p) = M^2 \phi(p)\ ,
\end{equation}
by substituting there the Fock
decomposition~(\ref{twobody}) of the state vector $\phi(p)$
(here and below we will omit, for shortness, all indices in the notation of the
state vector) and calculating the matrix elements
of the operator $\hat{P}^2$ in Fock space. After decomposition of the momentum operator in free and interacting part, we can easily
get the eigenstate equation~\cite{bckm}:
\begin{equation}\label{eq1b}
2(\omega\cd p)\int \tilde{H}^{int}(\omega\tau)\frac{d\tau}{2\pi}
\phi(p)= -\left[\left(\hat{P}^{(0)}\right)^2-M^{2}\right]\phi(p)\ ,
\end{equation}
where $\tilde H^{int}$ is the interaction Hamiltonian in momentum
space:
\begin{equation}
\label{hamG} \tilde{H}^{int}(\omega\tau)=\int
H^{int}(x)e^{-i(\omega\cd x)\tau}d^4x\ .
\end{equation}
According to the decomposition~(\ref{twobody}), the
conservation law for the momenta in each Fock component has the
form
\begin{equation} \label{k1n}
k_1+k_2+\cdots +k_{n}=p+\omega\tau_n\ .
\end{equation}
Hence, the action of the operator
$\left(\hat{P}^{(0)}\right)^2-M^2$ on the state vector reduces
to the multiplication of each Fock component by the
factor $(\sum_{l=1}^n k_l)^2-M^{2}=2(\omega\cd p)\tau_n$.
It is therefore convenient to introduce the notation
\begin{equation}
{\cal G}(p)=2(\omega\cd p)\hat{\tau}\phi(p)\ ,
\end{equation}
where $\hat \tau $ is the operator which, acting on a given
component $\phi_{n,\sigma\sigma' }$ of $\phi(p)$, gives $\tau_n
\phi_{n,\sigma\sigma' }$. ${\cal G}(p)$ has the Fock decomposition
which is obtained from Eq.~(\ref{twobody}) by the replacement of
the wave functions $\phi_{n,\sigma\sigma' }$ by the vertex
functions $\Gamma_{n }$ (which we will also refer to as the Fock
components) defined by
\begin{equation}\label{eq22}
\bar{u}_{\sigma'}(k_{1})\Gamma_{n}u_{\sigma}(p)=(s_{n}-M^2)\phi_{n,\sigma\sigma'}\ ,
\end{equation}
with $s_{n}=(k_{1}+\ldots k_{n})^2$.  The vertex function $\Gamma_n$ is represented graphically by the $n$-body function shown on Fig.(\ref{gamman}). 

Since for each Fock component
$s_{n}-M^2=2(\omega\cd p)\tau_{n}$, we can cast the eigenstate
equation in the form
\begin{equation}\label{eq3}
{\cal G}(p) = \frac{1}{2\pi}\int
\left[-\tilde{H}^{int}(\omega\tau)\right]\frac{d\tau}{\tau} {\cal G}(p)\ .
\end{equation}
The physical bound state mass $M$ is found from the condition that
the eigenvalue is 1.  This equation is quite general and
equivalent to the eigenstate equation $\hat P^2 \phi =M^2 \phi$.
It is nonperturbative in the sense that it sums up to all orders irreducible contributions which involve at most $N$ particles at any given (light-front) time.

\subsection {Eigenvalue equation}
According to our discussion in Sec.~\ref{general}, a test function should be attached to each particle creation or destruction operator in momentum space. 
It is then easy to extend the eigenvalue equation we developed in \cite{kms_04,kms_08} in order to account for these test functions. We can therefore apply the following rules:
\begin{itemize}
\item To each external boson or fermion line of momentum $k_i$, one should attach a factor $f({\bf k}_i^2/\Lambda^2)$.
\item To each internal propagator with momentum $k_j$, one should attach a factor $\left[ f({\bf k}_j^2/\Lambda^2)\right]^2$ .
\end{itemize}
Since each vertex function $\Gamma_n$ is attached to one fermion and $n-1$ boson lines (external or internal), it will be multiplied at least by $f({\bf k}_1^2/\Lambda^2) \ldots f({\bf k}_n^2/\Lambda^2)$. We can thus redefine $\Gamma_n$ to include implicitly these test functions. We shall call $\bar \Gamma_n$ these new vertex functions:
\begin{equation} \label{gammabar}
\bar \Gamma_n [k_1 \ldots k_n]=  \Gamma_n[k_1 \ldots k_n] f({\bf k}_1^2/\Lambda^2) \ldots f({\bf k}_n^2/\Lambda^2) \ .
\end{equation}
 In the limit where all the test functions go to $1$, one has $\bar \Gamma_n \to  \Gamma_n$. Since the $f's$ are SRTFs, this also implies that any $\bar \Gamma_n$ is also a super regular function with respect to all momenta. We will thus be able to apply the Lagrange formula to all loop calculations, and derive the extension of all singular distributions along the lines detailed in the previous section.
\begin{figure}[btph] 
\begin{center} 
\includegraphics[width=16pc]{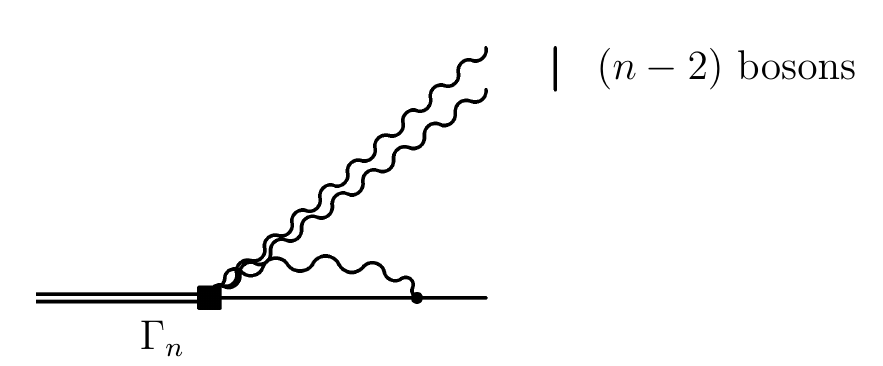}
\caption{Typical loop diagram entering the eigenvalue equation for the vertex function $\Gamma_n$.\label{self_gen}}
\end{center}
\end{figure}

Moreover, the only singular contributions are associated with the loop diagram of Fig.~(\ref{self_gen}) (or similar ones with contact interactions attached to the vertices). This generates both self-energy type diagrams to all orders (sometimes called rainbow diagrams) and overlapping loops to all orders. If one can give a mathematical sense to this contribution, one can give a mathematical sense to the eigenvalue equation to any order.
To show how this happens in practice, we shall consider in the following the Yukawa model in the two-body, $N=2$, approximation. 

\subsection{The Yukawa model in the $N=2$ approximation}
The Lagrangian we start from is given by
\begin{multline}
\label{yuka}
{\cal L}=\bar{\Psi}\left[i\sla{\partial}-m\right] \Psi +\frac{1}{2}\left[\partial_{\nu}\Phi
\partial^{\nu}\Phi-\mu^2\Phi^2\right] \\
+g\bar{\Psi}\Psi \Phi+\Delta m\bar{\Psi}\Psi \ ,
\end{multline}
where $\Delta m$ is the mass correction given by
\begin{equation} \label{deltam}
\Delta m = m - m_0\ .
\end{equation}
The fermion and boson fields are denoted by $\Psi$ and $\Phi$ respectively. In Eq.~(\ref{deltam}), $m$ denotes the (observable) physical mass of the fermion while $m_0$ is its (asymptotic) bare mass. The mass of the boson is $\mu$. Since in our formalism 
every amplitude is finite from the very beginning we do not need to consider any counterterm to cancel divergences. We introduce however 
a finite mass correction $\Delta m$ in (\ref{yuka}) in order to decompose our state vector on the basis of fields with physical mass m,
and treat the mass term $\Delta m \bar \Psi \Psi$ as an interaction. According to the general Fock sector dependent renormalization 
scheme detailed in \cite{kms_08}, the mass correction, as well as the bare coupling constant, should depend on the Fock sector under 
consideration, while the basis on which the Fock sectors are decomposed will stay the same.

From the Lagrangian (\ref{yuka}), we can easily deduce the effective Hamiltonian on the light-front. It is given in Ref.~\cite{kms_04}.
It includes contact interactions arising from the elimination of nondynamical degrees of freedom. Contrary to the Pauli-Villars 
regularization scheme, these contact interactions are not cancelled by Pauli-Villars fermions. 

As explained also in  \cite{kms_08}, 
we have to consider {\it a priori} a specific insertion on the light-front in order to correct for possible violation of rotational 
invariance. This  insertion is denoted by $Z_\omega$.

\begin{figure}[btph] 
\begin{center}
\includegraphics[width=20pc]{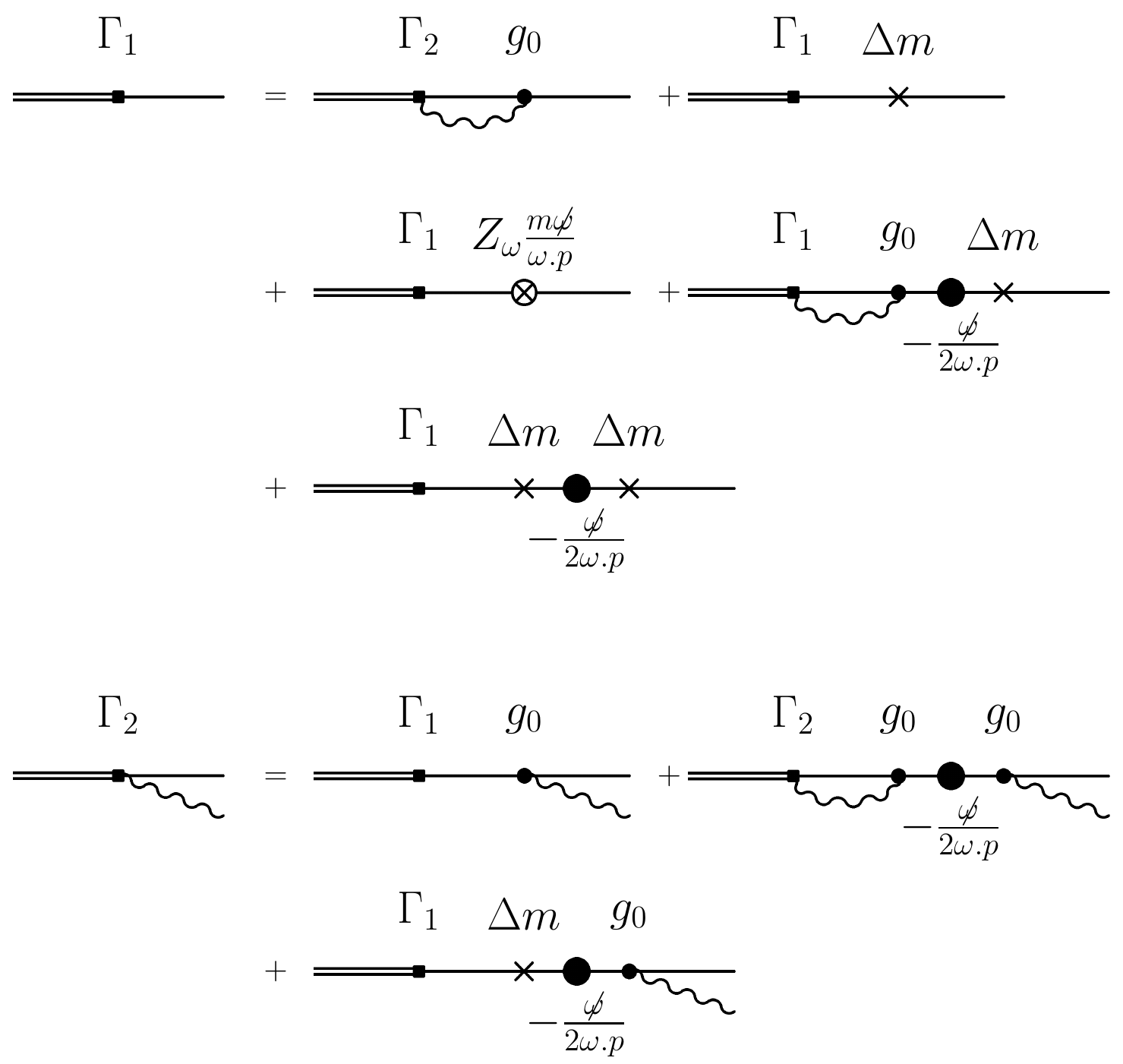}
\caption{Eigenvalue equation for the Yukawa model in the $N=2$ approximation\label{eigen}.}
\end{center}
\end{figure}

The eigenvalue equations are shown on Fig.~(\ref{eigen}). Since there is only one nonzero mass correction and bare coupling constant 
in this $N=2$ calculation, we denote them simply by $\Delta m$ and $g_0$ respectively, with no reference to the Fock sectors. 
The vertex functions are denoted by $\Gamma_1$ and $\Gamma_2$.

Following (\ref{Yuphi}), we can decompose the vertex functions in the following way:
\begin{subequations}
\begin{eqnarray}
\label{gh1}
\bar{u}(p_1)\bar \Gamma_1 u(p)&=&(m^2-M^2) \bar \varphi_1\\
\bar{u}(k_1)\bar \Gamma_2 u(p)&=&\bar{u}(k_1) \left[ \bar b_1+
\bar b_2\frac{m}{\omega\cd p}\sla{\omega}\right]u(p), \label{g2}
\end{eqnarray}
\end{subequations}
where $M$ is the bound state mass of momentum $p$, and $m$ is the mass of the fermionic constituent of momentum $p_1$ in the one-body component and $k_1$ in the two-body component. The functions $b_{1,2}$ are related to $\varphi_2, \varphi_2^\omega$ in (\ref{Yuphi}) using (\ref{eq22}). The corresponding quantities denoted with a bar are defined according to (\ref{gammabar}). The mass of the boson, of momentum $k_2$,  will be denoted by $\mu$.

To solve the eigenvalue equation shown on Fig.~(\ref{eigen}), we multiply both equations by $u$ on the left  and by $\bar u$ on the right, sum over spin indices, multiply by $1$ and $\sla \omega$ successively the second equation, and take the trace \cite{kms_04}. We thus get one equation $I_1=0$ from the first equation defining $\bar \Gamma_1$ and two equations $I_{2,3}=0$ from the equation defining $\bar \Gamma_2$. Solving the last two equations with respect to $\bar b_1$ and $\bar b_2$, we get for the calculation of the ground state properties for which $M \to m$
\begin{subequations}\label{b's}
\begin{eqnarray}
\bar b_1&=&2\ \bar \varphi_1\ g_0\ m\ f({\bf k}^2_1/\Lambda^2)\ f({\bf k}^2_2/\Lambda^2)\label{b1}\\
&&\nonumber \\
\bar b_2&=&-\bar \varphi_1\ g_0 \ \Delta m\  f({\bf k}^2_1/\Lambda^2)\ f({\bf k}^2_2/\Lambda^2)\label{b2} \\
&&-\frac{1}{2}g_0^2\frac{1}{(2\pi)^3} f({\bf k}_1^2/\Lambda^2)f({\bf k}_2^2/\Lambda^2)\nonumber \\
&&\int \frac{d^2k_\perp' dx'}{2x'(1-x')} \frac{\bar b'_1(2-x') + 2 \bar b'_2 (1-x')}{s'-m^2} \ , \nonumber
\end{eqnarray}
\end{subequations}
where $s'$ is the usual center of mass energy squared of  the internal boson - fermion state:
\begin{equation}
s'=\frac{{\bf k'_\perp}^2+m^2}{(1-x')}+\frac{{\bf k'_\perp}^2+\mu^2}{x'}\ .
\end{equation}
The usual longitudinal momentum fraction, with respect to the position $\omega$ of the light-front, is given by
 \begin{equation} 
 x=\frac{\omega \cd k_2}{\omega \cd p}\ . 
 \end{equation}
 The transverse momentum $k_\perp$ with respect to $\omega$ is defined from
 \begin{equation} \label{cinkt}
 R_2 = k_2 - x p\ \ \ \ \mbox{with} \ \ \ \  R_2 =  (R_2^0, {\bf k_\perp},R_2^0)\ ,
 \end{equation}
 since by construction $\omega \cd R_2 = 0$. The last component in (\ref{cinkt}) corresponds to the longitudinal momentum
 $R_2$ with respect to $\omega$. The prime quantities are defined similarly.

The vertex functions $\bar b_{1,2}$ do depend in principle on the kinematical variables $({\bf k_\perp},x)$ (and  $\bar b'_{1,2}$ on the variables  $({\bf k'_\perp},x')$) of the two-body system, as defined in Appendix (\ref{kin}). In the limit where $f \to 1$ however, these vertex functions should tend to constants according to Eqs.~(\ref{b's}). We say that they are almost constant. 

The limit $f\to 1$ can easily be done in Eq.(\ref{b1}), leading to
\begin{equation}
b_1=2\varphi_1g_0m\ .
\end{equation}

On top of the three equations $I_{1,2,3}$, we have to impose two normalization conditions \cite{kms_08}. The first one fixes the bare coupling constant as a function of the physical one, at the (nonphysical) on-energy shell point $s=m^2$
\begin{equation}
b_1(s=m^2) \equiv  g \sqrt{N_1} \ ,\label{g0g}
\end{equation}
where $N_1$ is the norm of the one-body state \footnote{Note a different definition of the bare coupling constant as compared to 
\cite{kms_08}, where the normalization correction is missing.}. This condition originates from the construction of the vertex function $\Gamma_2$ in 
which  the constituent fermion state (single solid line in Fig.~(\ref{gamman})) does not correspond to the fully dressed one-body state but
  to a state dressed with one less boson, due to the presence of an additional boson in flight at the same light-front time. This is a 
  direct consequence of the truncation of the Fock space. It should therefore be corrected by its normalization, i.e. by a factor 
  $\sqrt{1-N_2}=\sqrt{N_1}$ for a two-body Fock state truncation, where $N_2$ is the norm of the two-body state. 

The second renormalization condition insures that, always at $s=m^2$, i.e. on the energy shell, the two-body vertex function $\Gamma_2$ should not depend on the position of the light-front plane, i.e.
\begin{equation} \label{oes}
\bar b_2(s=m^2) \equiv 0 \ .
\end{equation}
With this latter condition, the normalization condition of the state vector writes \cite{kms_08} 
\begin{equation}
\label{norme}
1 = N_1+N_2=4m^2\varphi_1^2+4m^2\varphi_1^2 g_0^2 J_2\ ,
\end{equation}
with
\begin{multline}
\label{normJ2}
J_2 = \frac{1}{2(2\pi)^3}\int d^2k_\perp' dx'  x' \\
\frac{{\bf k'_\perp}^2+m^2(2-x')^2}{\left[{\bf k'_\perp}^2+m^2x'^2+\mu^2(1-x')\right]^2} f({\bf k}^2_1/\Lambda^2)f({\bf k}^2_2/\Lambda^2)\ .
\end{multline}
It is calculated in Appendix \ref{norma}. This fixes $\varphi_1$. The condition (\ref{g0g}) with (\ref{norme}) defines $g_0$:
\begin{equation} \label{radia}
g_0 =g
\end{equation}
and the norms of the one- and two-body states are 
\begin{subequations} \label{norms}
\begin{eqnarray}
N_1& = & \frac{1}{1+g^2 J_2}\ ,\\
N_2& = & \frac{g^2 J_2 }{1+g^2 J_2}\ .
\end{eqnarray}
\end{subequations}
The condition (\ref{radia}) insures that the bare coupling constant is not renormalized since, in the $N=2$ approximation, there is no radiative corrections to the vertices. The norms $N_1$ and $N_2$ in (\ref{norms}) are always less than $1$ and positive, but scale dependent.
Since $\bar b_2$ is almost constant and thus almost zero from (\ref{oes}), we immediately get from (\ref{b2}), 
\begin{multline}
\Delta m = - \frac{g}{4 \varphi_1}\frac{1}{(2\pi)^3}\\
\int d^2k'_\perp dx'\frac{(2-x')}{{\bf k'_\perp}^2+m^2x'^2+\mu^2(1-x')}\ \bar b_1'\ .
\label{dm}
\end{multline}
With this value for  $\Delta m$, the first equation $I_1=0$ defines the $\omega$-dependent insertion $Z_\omega$
\begin{multline}
Z_\omega = -\frac{g}{8 \varphi_1 m^2}\frac{1}{(2\pi)^3}\\
\int \frac{d^2k_\perp' dx' }{(1-x')}\frac{{\bf k'_\perp}^2+m^2 x'(2-x')}{{\bf k'_\perp}^2+m^2x'^2+\mu^2(1-x')}\ \bar b_1'\ .
\label{Zw}
\end{multline}
We have identified in (\ref{dm}) and (\ref{Zw}) $\bar \varphi_1$ with $\varphi_1$. Using the general properties of $\bar b_{1}$ as SRTFs, we can now calculate $\Delta m$ and  $Z_\omega$ as well as $J_2$ in (\ref{normJ2}) following the derivation outlined in Sec.(\ref{general}). This is done in Appendix~\ref{selfe}. This leads to 
\begin{eqnarray} \label{dm2}
\Delta m&=& -\frac{3 m g^2}{32\pi^2}\mbox{Log} [\eta]   +  \frac{m g^2}{16\pi^2}\\
&&\int_0^1dx' (2-x')\mbox{Log} \left[ \frac{m^2 x'^2+\mu^2(1-x')}{m^2}\right]  \nonumber\ ,\\
Z_\omega&=&0\ .  \label{zw}
\end{eqnarray}
These expressions for $\Delta m$ and $Z_\omega$ are  analogous to the ones found in \cite{kms_04}. They correspond to the usual calculation 
using PV regularization scheme - with the sum of different factors in  $\frac{\Lambda^2_{PV}}{m^2}$ taken in the limit $\Lambda^2_{PV}
\to \infty$ replaced by a specific function in the scaling parameter $\eta$ - but without any constraint on $\eta$ except that it 
should be larger than $1$.  The absence of an $\omega$-dependent counterterm, from $Z_\omega = 0$, is a clear check that the TLRS does not 
violate rotational invariance in LFD calculations, at least in the $N=2$ truncation.

In order to extend this approach to the more general case, it is necessary to calculate the contribution of Fig.~(\ref{self_gen}). 
This is a simple extension of the calculation of the self-energy done in this Section, apart from the presence of an external momentum
 $K= \sum_{i=3}^n k_i$, where the $k_i$'s are the momenta of the $(n-2)$ bosons in Fig.~Ê(\ref{self_gen}).

\section {Perspectives} \label{pers}
We consider in this study a coherent framework based on the construction of quantum fields as operator valued distributions acting on 
test functions, focusing on the specific properties the test functions should obey in order to achieve a generic quantum field theory 
description which preserves Poincar\'e and Lorentz invariance. These properties are of three different types:

{\it i)} In order to achieve the independence of any physical amplitude on the choice of test functions, and  to fulfill Poincar\'e 
and Lorentz invariance, we choose test functions as partitions of unity. 

{\it ii)} The generalization of the procedure to deal with singularity of any order, which is essential in nonperturbative calculations,
 requires one to choose partitions of unity which are superregular, i.e. which go to zero with all their derivatives at the boundaries. 

{\it iii)} Finally, the extension of distributions over the whole space should respect the inherent scaling behavior in the UV as well as IR domains. This is achieved by using test functions with a running boundary, in contrast to a fixed boundary similar to a naive cut-off. This scale invariance enables a renormalization group analysis of any physical observable. \\

This scheme has been applied to the calculation of a bound fermion-boson system in the Yukawa model within light-front dynamics. We show, 
 in this simple example, why this Taylor-Lagrange renormalization scheme is very natural to use in nonperturbative calculations on the
 light-front.  It is however quite general, and can be applied to any formulation of quantum field theory. 
 
 The mathematical properties of the test functions translate immediately to similar properties for the many-body vertex 
 function describing the Fock components of the state vector. The calculation of the anomalous magnetic moment of a fermion, 
 in the simplest two-body truncation, can be done very easily in the Yukawa model, following the derivation of Ref.~\cite{kms_08}, 
 but without any reference to Pauli-Villars fields and infinite mass  limit.

Our formulation can form the basis of more involved calculations including higher Fock state components, like for
instance the calculation of the anomalous magnetic moment of a fermion in a nontrivial three-body Fock state truncation. It can also be extended to the investigation of baryon properties at low energies in chiral effective field theory on the light-front  \cite{jfm}.

\section*{Acknowledgements}
We acknowledge financial support from CNRS/IN2P3 during the course of this work. E. Werner is grateful to Alain Falvard for his
 kind hospitality at the LPTA.


\appendix
\section{Construction of the test function} \label{PU}
\subsection{Partition of unity}
A partition of unity on the interval $[a,b]$ is built up from a family of functions $\beta_i(X)$ with
\begin{equation}
\sum_{i=1}^N \beta_i(X)=1 \ \ \ \mbox{for any}\ \ \ X \in [a,b]\ .
\end{equation}
A simple realization of this condition is to choose a basic function $u(X)$ such that
\begin{equation}
u(X)+u(h-X)=1 \ \ \ \mbox{for any}\ \ \ X \in [0,h]\ ,
\end{equation}
where $h$ is an arbitrary positive real number, and take
\begin{equation}
\beta_i(X) \equiv u(|X-ih|) \ \ \ \mbox{for} \ \ \ |X-ih| < h\ .
\end{equation}
\begin{figure}[btph] 
\begin{center} 
\includegraphics[width=15pc]{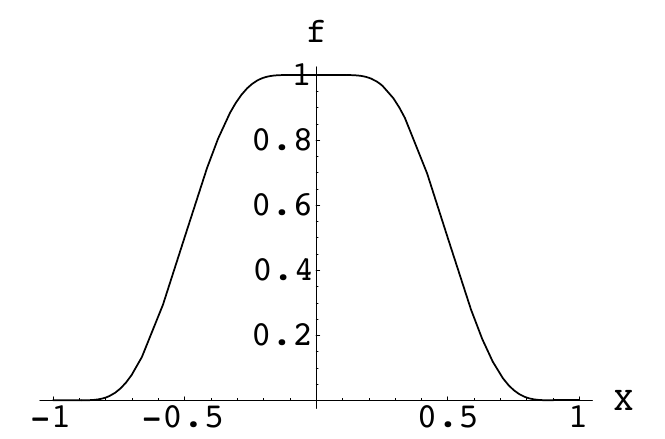}
\caption{Basic function $\beta_i(x)$ given in (\ref{betaj}) for $i=0,h=1$ and $\nu = 1$.\label{parua}}
\end{center}
\end{figure}
One typical example of such construction is shown on Figs.(\ref{parua},\ref{parub}). It corresponds to the following basic function
\begin{widetext}
\begin{equation}\label{betaj}
\beta_i(X) =\left\{
\begin{array}{lrl}
{\cal N} \int_{| X-ih |}^h dv \  \mbox{exp}\left[-\frac{h^{2\nu}}{v^\nu(h-v)^\nu} \right]  & \mbox{for} &  |X-ih| < h \\
0 & \mbox{for}& |X-ih| \ge h \ ,
\end{array} \right.
\end{equation}
\end{widetext}
with the normalization factor $\cal N$ given by
\begin{equation}
{\cal N}^{-1}=\int_{0}^h dv \  \mbox{exp}\left[-\frac{h^{2\nu}}{v^\nu(h-v)^\nu} \right] \ .
\end{equation}
\begin{figure}[btph] 
\begin{center} 
\includegraphics[width=15pc]{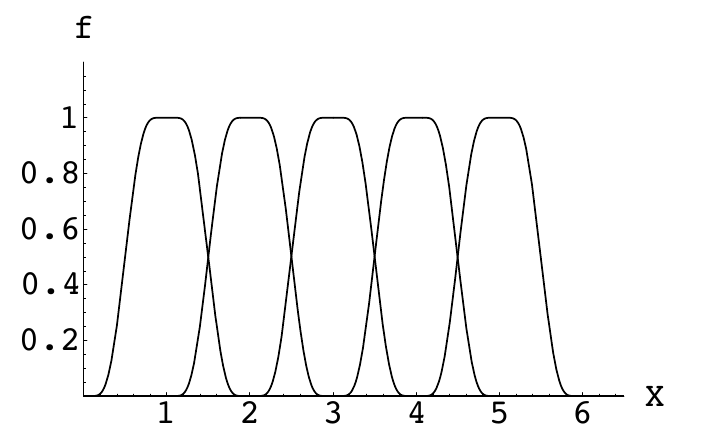}
\caption{Construction of a partition of unity from the basic functions $\beta_i(X)$ given in (\ref{betaj}), for $N=5$, $h=1$ and
 $\nu=1$. The partition of unity $f_{PU}(X)$ is the sum of all the basic functions.\label{parub}}
\end{center}
\end{figure}
The parameter $\nu$ is an arbitrary positive real number governing the shape of the function $\beta_i(X)$, from a triangle like shape to a square like one, with all its derivatives equal to zero at the boundaries given by $X=(i-1)h$ and $X=(i+1)h$. The function 
\begin{equation} \label{fpu}
f_{PU}(X)=\sum_{i=1}^N \beta_i(X)\ ,
\end{equation}
has thus the following properties
\begin{equation}
f_{PU}(X)= \left\{ \begin{array}{lrl}
u(h-X) & \mbox{for } & X \in [0, h]  \\
1  & \mbox{for } & X \in [h, Nh]  \\
u(X-Nh) & \mbox{for } & X \in [Nh, (N+1)h]\ . 
\end{array} \right.
\end{equation}

\subsection{Extension with a running support} \label{dyna}
The construction of the running test function introduced in Sec.\ref{DTF} can be done using the basic functions presented above.
 All the properties of the functions  $\beta_i$ are preserved when $h$ depends on $X$, with $h$ given by
 \begin{equation}
 h(X)=\eta^2 X g_\alpha (X)+ (\alpha-1) \ .
 \end{equation}
The additional constant $(\alpha-1)$ is chosen such that, in the IR domain,  the lower boundary of the support of the test function goes to zero when $\alpha$ goes to $1$. 

\begin{figure}[btph] 
\begin{center} 
\includegraphics[width=15pc]{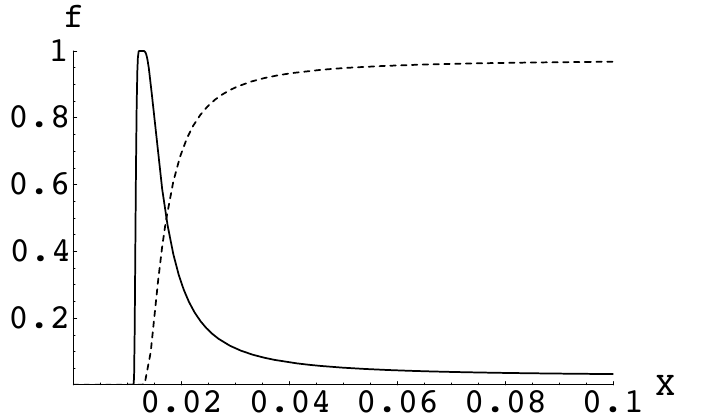}
\includegraphics[width=15pc]{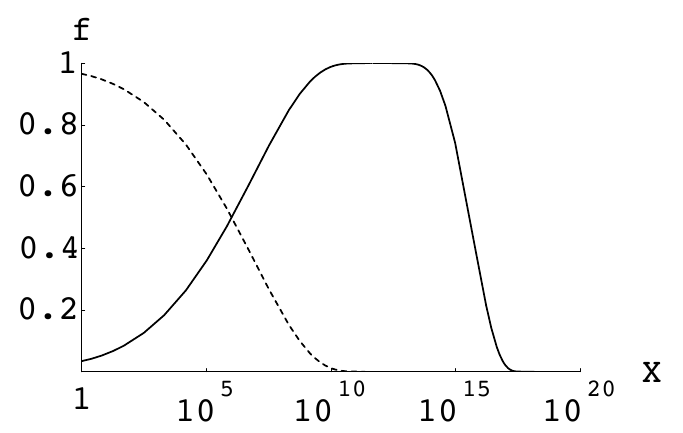}
\caption{Construction of a partition of unity with running support extension, for $i=0$ (dashed line) and $i=1$ (solid line), 
with $\alpha=0.95$ and  $\mu^2=2$. The upper curve shows the IR domain while the lower curve shows the UV domain. \label{parud}}
\end{center}
\end{figure}

With this choice of $h(X)$, it is sufficient to consider only two functions $\beta_i$, with $i=0,1$, so that the test function is given by
\begin{equation}
f(X;h)=\beta_0(X)+\beta_1(X)\ .
\end{equation}
The basic functions building up unity are shown on Fig.(\ref{parud}). 

This construction is of course not unique but is just mentioned here as an example of the construction of a running test function. 
We recall that the explicit form of the test function $f$ is indeed not necessary in the calculation of the physical amplitudes, 
as shown in Sec.~\ref{general}.

In the UV domain, i.e. for $X\to \infty$, the boundary of the test function, denoted by $H$,  is deduced from the boundary 
condition on $\beta_1$, i.e.  from $X-h=h$. We have thus
\begin{equation}
H(X)=2 h(X)=2 \eta^2 X g_\alpha(X)+2(\alpha-1)\ .
\end{equation}
For large X, and after redefinition of the arbitrary scale $2 \eta^2 \to \eta^2$, we have 
\begin{equation}
H(X)\simeq \eta^2 X^\alpha \ ,
\end{equation}
 which is precisely of  the form given in (\ref{running}). The maximal value of $X$, defined by $X_{max}=H(X_{max})$, is given by 
\begin{equation}
X_{max}=(\eta^2)^{\frac{1}{1-\alpha}}
\end{equation}
It goes to infinity when $\alpha \to 1^-$, with $\eta^2 >1$. The limit $f \to 1$ is thus achieved by the limit $\alpha \to 1^-$.
 
In the IR domain for any $X$ close to $0$, the test function is $0$ for $X<X_{min0}$ with $X_{min0}$ given by
\begin{equation}
X_{min0}=2h(X_{min0})=2\eta^2 X_{min0}^{\alpha}+2(\alpha-1)\ ,
\end{equation}
so that, for $\alpha \to 1^-$,
\begin{equation}
X_{min0}=\frac{2(1-\alpha)}{2\eta^2-1}\ .
\end{equation}
In turn the test function is $1$ for $X>X_{min1}$ with  $X_{min1}$ given by
\begin{equation}
X_{min1}=h(X_{min1})=\eta^2 X_{min1}^{\alpha}+(\alpha-1)\ ,
\end{equation}
so that, for $\alpha \to 1^-$,
\begin{equation}
X_{min1}=\frac{(1-\alpha)}{\eta^2-1}\ .
\end{equation}
 It also goes to zero like $(1-\alpha)$. Hence when $\alpha \to 1^-$ the test function effectively tends to a step function. In  this region where 
 the test function is less than $1$, we have $(1-\alpha)X > h((1-\alpha)X)$. This defines the lower limit for the $t$ integration in Eq.~(\ref{TIR1}), 
 with $(1-\alpha)X/t > h((1-\alpha)X/t)$  i.e.
\begin{equation}
t>X\left(\eta^2-1\right)\ .
\end{equation}
Since $\eta^2$ is an arbitrary real number larger than $1$, we can define $\tilde \eta = \eta^2-1$, with $\tilde \eta$ an 
arbitrary positive real number. We have in that case  $t>\tilde \eta X$.

\section {Calculation of the vertex functions} \label{selfe}
\subsection{Kinematics} \label{kin}
It is more appropriate to calculate all momenta in the reference frame where ${\bf p}=0$ since this reference frame is the same whatever the
 number of constituents we have in the Fock sector. For the calculation of the self-energy type diagram, we denote by  $k_1$ ($k_2$) 
 the momentum of the intermediate fermion (boson). They satisfy the on-mass shell conditions, and their vector parts are decomposed in
  parallel $k_\parallel$ and perpendicular  $k_\perp$ components, relative to the position of the light-front $\omega = (\omega_0,\vec{\omega
  })$. We have therefore, with $p^2=M^2$
\begin{subequations}
\begin{eqnarray}
 \omega \cd p &=& \omega_0 M \ ,\\
  \omega \cd k_2 &= &x\ \omega _0 M= \omega _0 (\varepsilon _{k_1} - k_{1,\parallel})\ ,
\end{eqnarray}
\end{subequations}
with  $\varepsilon_{{\bf k}_1}=\sqrt{{\bf k}_{\perp}^2+k_{2,\parallel}^2+\mu^2}$. We thus get
$$
k_{2,\parallel}=\varepsilon_{k_1} - x M\ ,
$$
and
$$
 k_{2,\parallel}=\frac{{\bf k}_{\perp}^2+\mu^2-x^2M^2}{2xM},\,\,\,\,\,\,\,\,
\varepsilon_{{\bf k}_2}=\frac{{\bf k}_{\perp}^2+\mu^2+x^2M^2}{2xM}\ ,
$$
so that
\begin{eqnarray}
{\bf k}_2^2 &=&\varepsilon_{k_2}^2-\mu^2  \nonumber \\
&=&\left( \frac{{\bf k}_\perp^2+\mu^2}{2xM}+\frac{xM}{2}\right)^2-\mu^2\ .
\label{k2}
\end{eqnarray}
In the limit of large transverse momentum, or for $x\to 0$, we have
\begin{equation} \label{k2inf}
{\bf k}_2^2 \approx  \frac{({\bf k}_\perp^2+\mu^2)^2}{4x^2M^2}\ .
\end{equation}
The momentum ${\bf k}_1$ of the intermediate fermion can be calculated easily with
the replacement $x \to 1-x$, $m \to \mu$ and ${\bf k}_\perp \to -{\bf k}_\perp$,
so that
\begin{equation} \label{q2}
{\bf k}_1^2=\left( \frac{{\bf k}_\perp^2+m^2}{2(1-x)M}+\frac{(1-x)M}{2}\right)^2-m^2 \ .
\end{equation}
In the limit of large transverse momentum, or for $x \to 1$, we have
\begin{equation} \label{k1inf}
{\bf k}_1^2 \approx \frac{({\bf k}_\perp^2+m^2)^2}{4(1-x)^2M^2}\ .
\end{equation}

\subsection {The mass correction} \label{apb2}
The equations for the two-body components lead to the expression (\ref{dm}) for the mass correction. Using (\ref{b1}), it is enough to consider the following integral
\begin{equation}
I= \int_0^1dx' \int_0^\infty d^2 k'_\perp  \frac{f({\bf k'}_1^2/\Lambda^2) f({\bf k'}_2^2/\Lambda^2)}{{\bf k'}_\perp^2+m^2x'^2+\mu^2 (1-x')} \ ,
\end{equation}
where ${\bf k'}_1^2 $ and ${\bf k'}_2^2 $ are given in  Eqs.~(\ref{k2},\ref{q2}),  in terms of $x'$ and ${\bf k}'_\perp$, and for the most dangerous region in ${\bf k'}_\perp^2$ (${\bf k'}_\perp^2 \to \infty$) and $x'$ ($x' \to 0,1$) by (\ref{k2inf}) and (\ref{k1inf}).

Since $I$ has no singularities in $x'=0,1$, we can even consider:
\begin{eqnarray}
{\bf k'}_1^2&\to& \frac{{\bf k'}_\perp^4}{4(1-x')^2m^2}\ , \\
{\bf k'}_2^2&\to &\frac{{\bf k'}_\perp^4}{4x'^2m^2}\ .
\end{eqnarray}
The test functions will alternatively correct the behavior of the integrand of $I$ for $x'\sim 0$, with f(${\bf k'}_2^2$)
 or for $x'\sim 1$, with f(${\bf k'}_1^2$). We can thus divide $I$ in two contributions $I_{1,2}$ for $x'<1/2$ and $x'>1/2$. 
 Keeping only the test  function which is not identically $1$ in each interval, we have, for $I_1$ for instance:
\begin{equation}
I_1= \int_0^{\frac{1}{2}}dx'\int_0^\infty d^2 k'_\perp
\frac{f({\bf k'}_\perp^4/4x'^2m^2\Lambda^2)}{{\bf k'_\perp}^2+m^2 x'^2+\mu^2 (1-x')} \ ,
\end{equation}
where the argument of the test function is valid only in the ${\bf k'}_\perp^2 \to \infty$ region.
With the following change of variable 
\begin{equation} \label{chX}
X={\bf k'}_\perp^2/2 m \Lambda x' \  ,
\end{equation}
we have 
\begin{multline}
I_1=  2 \pi m \Lambda \int_0^{\frac{1}{2}}dx'  x'  \\
\int_0^\infty dX \frac{f(X^2)}{2 m \Lambda x' X+m^2 x'^2+\mu^2 (1-x') } \ .
\end{multline}
To get rid of any momentum scale in the behavior of the integrand, we make the change of variable 
\begin{equation} \label{chY}
Z=\frac{m^2 X}{m^2x'^2+\mu^2(1-x')}\ ,
\end{equation}
so that $I_1$ writes
\begin{multline}
I_1= 2\pi \frac{\Lambda}{m} \int_0^{\frac{1}{2}}dx'   x'   \\
\int_0^\infty dZ \frac{f(\frac{Z^2}{m^4} [m^2x'^2+\mu^2(1-x')]^2)}{\frac{2\Lambda x'}{m} Z+1} \ .
\end{multline}
Using (\ref{afind}) with (\ref{la3a}) and  (\ref{La22}), for $k=0$, we have

\begin{multline}
I_1= 2\pi \frac{\Lambda }{m} \int_0^{\frac{1}{2}}dx'  x' 
\int_0^\infty dZ \partial_Z \left[\frac{Z}{\frac{2\Lambda x'}{m} Z+1}\right] \\
\int_a^\infty \frac{dt}{t}F\left(Z^2t^2, Z^2\right)\ ,
\end{multline}
with
\begin{equation} \label{atrans}
a=\frac{m^2x'^2+\mu^2(1-x')}{m^2}
\end{equation}
Because of  the structure of the test function, the integration domain in $t$ is limited according to the support of $F$, i.e. 
\begin{equation}
Z^2t^2 \le H(Z^2) \approx \eta^2 Z^2g_\alpha(Z^2) \ .\label{mint}
\end{equation}
Moreover, the integrand for the $Z$-integration behaves now as $1/Z^2$ when $Z \to \infty$, i.e. the integration over $Z$ is convergent. 
We may thus put $f\to 1$ (by letting $\alpha$ go to $1^-$) and the only remaining trace of the test function is in the scale $\eta$.
Doing the integration over $Z$, it  remains
\begin{eqnarray}
I_1&=& \pi \int_0^{\frac{1}{2}}dx' \int_a^\eta \frac{dt}{t} \nonumber \\
&=&\pi \int_0^{\frac{1}{2}}dx'\  \mbox{Log}\left[\frac{\eta \ m^2}{m^2x'^2+\mu^2(1-x')}\right] \nonumber \ .
\end{eqnarray}
We can do the same calculation for the second part $I_2$ of $I$ and get a similar result, the only difference being the integration 
limit on $x'$ from $1/2$ to 1. We thus get for $I$:
\begin{equation}
I =\pi \  \mbox{Log}[\eta]  
- \pi  \int_0^1 dx' \ \mbox{Log}\left[\frac{m^2x'^2+\mu^2(1-x')}{m^2}\right]\ .
\end{equation}
This gives the result indicated in Eq.~(\ref{dm2}).
\subsection {The $\omega$-dependent insertion} \label{apz}
The equation for the one-body component leads to the expression (\ref{Zw}) for the $\omega$-dependent  insertion.
It is enough to consider the integral
\begin{multline}
{\cal J}= \int_0^1dx \int_0^\infty d({\bf k'}^2_\perp) f({\bf k'_1}^2/\Lambda^2) f({\bf k'_2}^2/\Lambda^2)\\
\frac{{\bf k'_\perp}^2+m^2 x(2-x)}{(1-x)[{\bf k'_\perp}^2+m^2 x^2+\mu^2 (1-x)]} \ .
\end{multline}
${\cal J}$ can be expressed as
\begin{widetext}
\begin{eqnarray}
{\cal J}&=&\int_0^1\frac{dx}{1-x} \int_0^\infty d({\bf k'}^2_\perp) f({\bf k'_1}^2/\Lambda^2) f({\bf k'_2}^2/\Lambda^2)
+\int_0^1dx \int_0^\infty d( {\bf k'}^2_\perp)  (2 m^2x -\mu^2)
\frac{f({\bf k'_1}^2/\Lambda^2) f({\bf k'_2}^2/\Lambda^2)}{{\bf k'_\perp}^2+m^2 x^2+\mu^2 (1-x)} \nonumber \\
&\equiv&{\cal J}_1+{\cal J}_2 \ .
\end{eqnarray}
\end{widetext}
${\cal J}_1$ is well defined since the test function takes care explicitly of the behavior at high transverse momenta and at $x=1$  or at $x=0$ after the change of variable $x \to 1-x$.
With the asymptotic expressions at large ${\bf k'_\perp}^2$ given by (\ref{k2inf}), (\ref{k1inf}) and after an evident change of variable 
\begin{equation} \label{J1}
{\cal J}_1=2m\Lambda \int_0^1\frac{dx}{x} \int_{\frac{m}{2\Lambda}}^{\infty}dX \\ 
f\left[\frac{X^2}{x^2}\right]  f\left[\frac{(X+\frac{\mu^2-m^2}{2 m \Lambda})^2}{(1-x)^2}\right] \ .
\end{equation} 
The first test function treats the singularity at $x=0$. It is sufficient to apply the extension formula (\ref{TIR1}) for k=0. With $T(x) = 1/x$, 
we have
\begin{eqnarray}
\tilde T^<(x)&=&-\partial_x\left[x\int_{\tilde \eta x}^1 \frac{dt}{t^2}\ \frac{t}{x}\right]\nonumber\\
&=&\partial_x \mbox{Log} [\tilde \eta x]\nonumber
\end{eqnarray}
where $\tilde \eta$ is an arbitrary (dimensionless) scale. The derivative of the Log function should be understood in the sense of 
distributions:
\begin{multline}
<\frac{d}{dx}\mbox{Log}(x),\varphi>=\mbox{Lim}_{\epsilon \to 0} \int_\epsilon^\infty \mbox{Log}(x)\left[ -\varphi'(x)\right] dx \\
=\mbox{Lim}_{\epsilon \to 0}\left[\varphi(\epsilon)\mbox{Log}(\epsilon) +\int_\epsilon^\infty \frac{\varphi(x)}{x} dx \right] \ .
\end{multline}
With $\varphi(\epsilon)=\varphi(0)+\epsilon \varphi'(\xi)$, and since $\varphi'(\xi)$ is finite, we have
\begin{multline}
<\frac{d}{dx}\mbox{Log}(x),\varphi>=\\
\mbox{Lim}_{\epsilon \to 0} \left[ \int_\epsilon^\infty \frac{\varphi(x)}{x} dx + \varphi(0) \mbox{Log}(\epsilon) \right]\ .
\end{multline}
This is precisely the definition of the pseudofunction, denoted by Pf, of $1/x$ introduced in \cite{schwartz}, so that 
\begin{equation}
\tilde T^<(x)\equiv Pf\left[ \frac{1}{x}\right] \ .
\end{equation}
Hence ${\cal J}_1$  writes, after the change of variable $\frac{1}{W}=\frac{X+\frac{\mu^2-m^2}{2 m \Lambda}}{1-x}$,
\begin{eqnarray}
{\cal J}_1&=&2m\Lambda \int_0^1 dx Pf(\frac{1-x}{x}) \int_{0}^{\frac{2 m \Lambda(1-x)}{\mu^2}} \frac{dW }{W^2}
f(\frac{1}{W^2}) \nonumber \\
&=&2m\Lambda \int_0^1 dx (1-x)Pf(\frac{1}{x}) \left.(-\frac{1}{W})\right|^{W=\frac{2 m \Lambda(1-x)}{\mu^2}} \nonumber \\ 
&=&- \mu^2 \left.\mbox{Log}(x)\right|^{x=1} = 0,\nonumber
\end{eqnarray}
since the remaining test function defines $Pf(\frac{1}{W^2})$ and its integral outside the singularity is just $-\frac{1}{W}$. The notation $f(u)\vert ^{u=a}$ indicates simply that $f(u)$ should be taken at the value $u=a$, the lower limit of integration being taken care of by the definition of the pseudofunction.

We can now proceed to the calculation of ${\cal J}_2$. With the same change of variable in the ${\bf k'}^2_\perp$-integration 
leading to (\ref{J1}), we have
\begin{eqnarray} \label{J2}
{\cal J}_2&=&\int_0^1 dx \int_{\frac{m}{2\Lambda}}^{\infty}dX f\left[\frac{X^2}{x^2}\right]
f\left[\frac{(X+\frac{\mu^2-m^2}{2m\Lambda})^2}{(1-x)^2}\right]  \nonumber \\
& &\frac{2m^2(1-x)-\mu^2}{X+\frac{\mu^2}{2m\Lambda}x-\frac{m^2}{2\Lambda}x(2-x)} \nonumber \\
&=&\int_0^1 dx \int_{0}^{1}\frac{dY}{Y} f\left[\frac{m^2}{4Y^2\Lambda^2(1-x)^2}\right]  \nonumber \\
& &f^2\left[\frac{ m^2}{4Y^2\Lambda^2x^2}\right]\frac{2m^2(1-x)-\mu^2}{1+Y x (\mu^2-m^2(2-x))}.\nonumber
\end{eqnarray}
The arguments of the test functions are here taken for $Y$ close to $0$ since this is the domain where they differ from $1$. 

The product of the two test functions being invariant under the change $x \to 1-x$,  it  can  be represented by a single function
of argument $x(1-x)$ with the same support, say $G\left[\frac{A^2}{Y^2 x^2(1-x)^2}\right]$. With $Z(x)=\frac{A}{ x(1-x)}$, the Lagrange formula 
(\ref{la3IR}) applied on $G$ gives
\begin{eqnarray}
G\left[Z^2(x)\right]&=&-\int_{1}^{\infty}dt\partial_{t}G\left[Z^2(x) t^2\right] \nonumber \\
&=&\frac{x(1-x)}{1-2x}\partial_x\!\!\int_{1}^{\infty}\!\!\frac{dt}{t} G\left[Z^2(x)t^2\right] \nonumber.
\end{eqnarray}
Integrating by part on $x$,  ${\cal J}_2$ finally writes
\begin{eqnarray}
&&{\cal J}_2=\int_{0}^{1}\!\!\frac{dY}{Y}\int_0^1\!\!dx\int_{1}^{\infty}\!\!\frac{dt}{t} G\left[\frac{A^2 t^2}{Y^2 x^2(1-x)^2}\right]
\nonumber \\
&&\partial_{x} \left[\frac{x(1-x)(2m^2(1-x)-\mu^2)}{(1-2x)(1+Yx(\mu^2-m^2(2-x)))}\right]\ . \nonumber
\end{eqnarray}
The test function $G$ treats the singularity at $Y=0$ by the extension of $\frac{1}{Y}$ to $Pf(\frac{1}{Y})$.  Thereupon its support can be
extended to infinity, the upper integration limit in $t$ is restricted to a constant $\eta$  and $G$ becomes unity on every domain 
of integration. The remaining integration on $x$ is then trivially zero.

\subsection{Normalization of the state vector} \label{norma}
The calculation of (\ref{normJ2}) is very similar to the calculation of the mass correction $\Delta m$ detailed in the previous subsection.  We can simply rewrite $J_2$ as
\begin{widetext}
\begin{equation}
J_2= (4m^2-\mu^2)\int_0^1dx' \int_0^\infty d^2 k'_\perp  \frac{1-x'}{\left[{\bf k'_\perp}^2+m^2x'^2+\mu^2 (1-x')\right]^2}
+ \int_0^1dx' \int_0^\infty d^2 k'_\perp  \frac{f({\bf k'_1}^2/\Lambda^2) f({\bf k'_2}^2/\Lambda^2)}{{\bf k'_\perp}^2+m^2x'^2+\mu^2 (1-x')} \ ,\nonumber
\end{equation}
\end{widetext}
The first term is convergent (hence the limit $f\to 1$ taken), while the second one is exactly $I$ calculated in (\ref{apb2}). 
It depends logarithmically on $\eta^2$. Its explicit expression is not needed in our calculation.
%
 

\end{document}